\documentclass[
    ,final            
  ]
  {aipproc}

\layoutstyle{6x9}

\begin{document}

\title{Probing internucleus potential with large-angle 
quasi-elastic scattering}


\author{K. Hagino}{
address={Department of Physics, Tohoku University,
Sendai 980-8578, Japan}
}

\author{K. Washiyama}{
address={Department of Physics, Tohoku University,
Sendai 980-8578, Japan}
}

\begin{abstract} 
Recent measurements for fusion cross section 
at energies around the Coulomb barrier 
have systematically indicated 
a significant deviation of fusion cross sections from a prediction of 
double-folding model. It has been argued that the deviation can be 
accounted for if one uses a large value of surface diffuseness parameter 
for a Woods-Saxon internuclear potential. 
We investigate this problem 
using large-angle quasi-elastic scattering, that is a good counterpart 
of fusion reaction. 
Towards a reconciliation of the apparent 
anomaly in the diffuseness parameter for fusion reactions, 
we also discuss possible ingredients which are missing in present 
nuclear reaction models. 
Those include a deviation from the frozen density approximation, 
the effect of 
antisymmetrization and the role of coordinate dependent 
moment of inertia for the relative motion between the 
colliding nuclei. 
\end{abstract}

\maketitle


\section{Introduction}

It has been recognized for some time that fusion reactions at 
energies around the Coulomb barrier require a large value of 
surface diffuseness parameter if one uses the Woods-Saxon parametrization 
for an internuclear potential
[1--8].
For calculations of 
elastic and inelastic scattering,
which are sensitive only to the surface region of 
the nuclear potential,  
the diffuseness parameter 
of around 0.63 fm has been conventionally employed 
\cite{broglia91,Christensen76}.
This value of surface diffuseness parameter has been well accepted, 
partly because it is consistent with 
a double folding potential \cite{SL79}. 
In contrast, 
a recent systematic study has shown that
experimental data for heavy-ion fusion reactions 
require a larger value of the diffuseness parameter, 
ranging between 0.75 and 1.5 fm, 
as long as the Woods-Saxon 
parameterization is employed \cite{newton04}.

In this contribution, we present our recent systematic analyses 
on this problem using 
large-angle quasi-elastic scattering \cite{WHD06,hagino05}. 
Quasi-elastic scattering (a sum of elastic, inelastic, and transfer 
channels) is a good counterpart of fusion 
reaction \cite{hagino04,ARN88}. 
Firstly, both are inclusive processes in a sense that the final configurations 
(channels) of the projectile and target nuclei are all summed up. 
The former is related to the reflection probability at the Coulomb barrier, 
while the latter 
to the penetration probability. 
Since the reflection and the penetration probabilities are related to 
each other by the unitarity condition, so are the quasi-elastic 
and fusion cross sections. 
Secondly, at energies close to the Coulomb barrier, 
both are sensitive to 
the collective inelastic excitations of the colliding nuclei
and/or transfer process \cite{nanda98,BT98}. 
Using this fact, the experimental barrier distribution, 
originating from 
the channel coupling effects\cite{rowley91}, 
has been extracted for many systems from both fusion and quasi-elastic 
cross sections \cite{L95,nanda98,timmers95}. 

\section{Large-angle quasi-elastic scattering at deep sub-barrier energies}

\subsection{Advantages of using deep sub-barrier data}

In investigating internuclear potentials using heavy-ion 
quasi-elastic scattering, 
we are particularly interested in the deep sub-barrier region. 
At these energies, 
the cross sections of (quasi)elastic scattering are close to the 
Rutherford cross sections, with small deviations 
caused by the effect of nuclear interaction. 
This effect can be taken into account 
by the semiclassical perturbation theory. 
The ratio of elastic scattering $\sigma_{\rm el}$
to Rutherford cross sections $\sigma_R$ at a backward angle $\theta$
is given by \cite{hagino04,LW81}
\begin{equation}
\frac{d\sigma_{\rm el}(E_{cm},\theta)}{d\sigma_R(E_{cm},\theta)}
\sim
1+\frac{V_N(r_c)}{ka}\,
\frac{\sqrt{2a\pi k\eta}}{E_{cm}}, 
\label{deviation}
\end{equation}
where 
$E_{cm}$ is the center-of-mass energy,
$k=\sqrt{2\mu E_{cm}/\hbar^2}$, $\mu$ being the reduced mass, 
and $\eta$ is the Sommerfeld parameter. 
This formula is obtained by assuming 
that the nuclear potential $V_N(r)$ has an exponential form,
$\exp(-r/a)$,
around the distance of closest approach, $r_c$. 
We see from this formula
that the deviation of the elastic cross sections
from the Rutherford ones is sensitive to the surface region 
of the nuclear potential, especially to the surface diffuseness 
parameter $a$. 

\begin{figure}
  \includegraphics[height=.45\textheight]{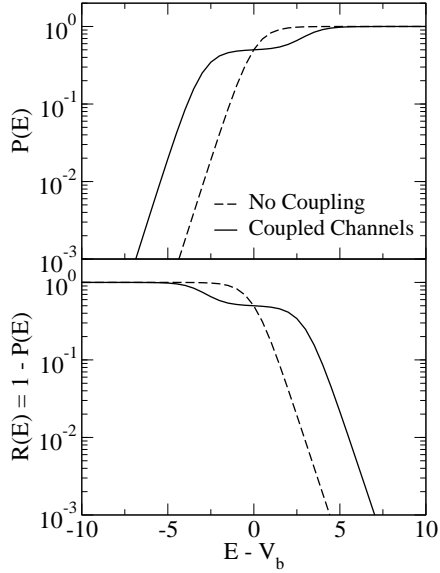}
  \caption{The penetrability (the upper panel) and the reflection 
probability (the lower panel) for a two-channel problem. 
}
\end{figure}

There is another advantage of using the deep sub-barrier 
data. 
That is, 
the effect of channel coupling on 
quasi-elastic scattering can be disregarded 
at these energies, since the reflection 
probability is almost unity irrespective of 
the presence of channel couplings, even though inelastic 
channels themselves may be strongly populated \cite{hagino05}. 
This is similar to fusion at energies well above the Coulomb barrier, 
where the penetrability is almost unity \cite{newton04}. 
We illustrate this in Fig. 1 for a two-channel problem. 

From these considerations, it is evident that 
the effect of surface diffuseness parameter can be studied 
in a transparent and unambiguous way using the large-angle quasi-elastic
scattering at deep sub-barrier energies. 

\subsection{Spherical systems} 

Let us now show the results of our analyses. 
In order to analyze the experimental data 
at deep sub-barrier energies,
we use a one-dimensional optical potential 
with the Woods-Saxon form. 
Absorption following transmission through the barrier
is simulated by an imaginary potential that is 
well localized inside the Coulomb barrier.
This model calculates the elastic and 
fusion cross sections, in which the elastic cross sections can be 
considered as quasi-elastic cross sections to a good approximation 
at these deep sub-barrier energies. 

In order to carry out a systematic study,
we estimate the Coulomb barrier height using 
the Aky\"uz-Winther potential \cite{broglia91}. 
We then vary the surface diffuseness parameter while keeping the 
barrier height. 
We define the region of ``deep sub-barrier energies''
as the region 
where 
the experimental value of the ratio of the quasi-elastic 
to the Rutherford cross sections
is larger than around 0.94.
See Ref. \cite{WHD06} for more details. 

Figure 2 compares the experimental data with
the calculated cross sections obtained with 
different values of the surface diffuseness 
parameter 
for the $^{32}$S + $^{197}$Au system (the upper panel) 
and the $^{34}$S + $^{197}$Au system (the lower panel). 
The best fitted values for the surface diffuseness parameter 
are $a=0.57\pm 0.04$ fm and $a=0.53\pm 0.03$ fm 
for the $^{32}$S and $^{34}$S + $^{197}$Au reactions, respectively. 
The cross sections obtained with these surface diffuseness 
parameters are denoted by the solid line in the figure. 
The dotted and the dot-dashed lines are calculated with
the diffuseness parameter of $a$ = 0.80 fm and $a$ = 1.00 fm,
respectively.
It is evident from the figure that 
these spherical systems 
favor the standard value of the surface diffuseness parameter,
around $a=$ 0.60 fm. 
The calculations with the larger diffuseness 
parameters underestimate the quasi-elastic cross sections 
and are not consistent with the energy dependence of the 
experimental data.
We obtain a similar conclusion for the 
$^{32,34}$S + $^{208}$Pb and 
$^{16}$O + $^{208}$Pb systems\cite{WHD06}. 

\begin{figure}
  \includegraphics[height=.4\textheight]{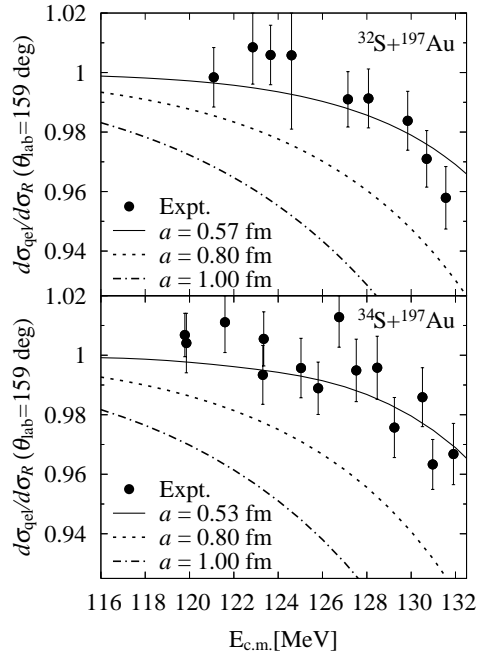}
  \caption{
The ratio of the quasi-elastic to the Rutherford cross sections
at $\theta_{\rm lab}=159^\circ$  
for the $^{32}$S + $^{197}$Au (the upper panel) reaction
and for the $^{34}$S + $^{197}$Au (the lower panel) reaction.
}
\end{figure}

\subsection{Deformed systems} 

We next analyse deformed systems, 
$^{16}$O + $^{154}$Sm, $^{186}$W \cite{WHD06,hagino05}. 
For these systems, only a few data points are available at
deep sub-barrier energies.
We therefore include 
the experimental data at energies not only well below but also 
around the lowest barrier  
in the $\chi^2$ fitting procedure. 
At these energies, 
the channel coupling effects start playing an important role in 
quasi-elastic reactions, 
and we 
include the effect of deformation of the target nucleus 
in our calculations \cite{WHD06,hagino04,ARN88}. 

The best fitted value for the surface diffuseness 
parameter obtained in this way is $a=1.14 \pm 0.03$ fm and 
0.79 $\pm 0.04$ fm for the $^{16}$O + $^{154}$Sm 
and $^{16}$O + $^{186}$W reactions, respectively. 
For the $^{16}$O + $^{154}$Sm reaction, the calculated cross sections 
with the standard value of the surface diffuseness parameter 
around 0.60 fm are found to be in strongly 
disagreement with the experimental data.

Although these large values of surface diffuseness parameter are 
consistent with that extracted from 
fusion, the origin of the difference between the spherical and the 
deformed systems is not clear. 
One should bear in mind, however, that 
our analyses for the deformed systems are somewhat model dependent as 
we needed to include 
the deformation effect in the calculations. 
Also, there might be a problem of normalization in the experimental 
data \cite{HD06}. 
Since these experiments were performed in aiming at extracting the 
quasi-elastic barrier distribution, rather than extracting the 
value of surface diffuseness parameter, 
the experimental data were arbitrarily normalized to unity 
at the lowest energy. 
(Notice that the shape of barrier distribution is rather insensitive to 
the absolute normalization factor.)
If the data had been normalized at a different energy, 
a different conclusion concerning the value of surface diffuseness 
parameter might have been achieved. 

In order to clarify the difference in the diffuseness parameter
between the spherical and the deformed systems, 
apparently 
further precision measurements for large-angle quasi-elastic scattering 
at deep sub-barrier energies will be necessary, especially for
deformed systems. 

\section{Discussions}

The value of the surface diffuseness parameter obtained in this 
study for the spherical systems agrees well with the 
conventionally used value $a\sim$ 0.63 fm. 
This value is consistent with the double folding potential, suggesting 
that the double folding model is valid at least 
in the surface region and for systems which do not involve 
a deformed target. 
The discrepancy between the values 
of the diffuseness parameter determined from fusion data 
and those from quasi-elastic data 
must be related with the dynamics inside the Coulomb barrier 
\cite{newton04}.

In this connection, it is instructive to step back to the original 
many-body Hamiltonian of the system, 
\begin{equation}
H=\sum_it_i+\sum_{i<j}v_{ij},
\label{H}
\end{equation}
where $t_i$ is the kinetic energy for a nucleon $i$ and $v_{ij}$ is a 
two-body interaction between nucleons. The philosophy of any reaction model 
is to somehow extract the degree of freedom for the 
relative motion from the original 
many-body Hamiltonian (\ref{H}) and write it in a form of 
\begin{equation}
H=H_{\rm rel}+H_{\rm s.p.}+H_{\rm coup},
\label{LACM}
\end{equation}
where $H_{\rm rel}$ is the Hamiltonian for the relative motion which 
describes the scattering process (the ``collective'' degree of freedom), 
$H_{\rm s.p.}$ is the Hamiltonian 
for other degrees of freedom than the relative motion
(``non-collective''  
degrees of freedom), and $H_{\rm coup}$ 
is the coupling term between the relative motion and the non-collective 
degrees of freedom. 

Although it is an extremely difficult task to extract the degree of freedom 
for the relative 
motion from the original many-body Hamiltonian in a 
consistent manner, let us assume that it can be achieved in some way.  
Yet, it is difficult to 
treat the coupling term, $H_{\rm coup}$ exactly. 
One usually takes one of the following approaches for this purpose. 
\begin{itemize}
\item Sudden approach

This approach is justified if the reaction takes place suddenly and/or 
the coupling term $H_{\rm coup}$ is negligibly small. 
The double folding model\cite{broglia91}, 
the optical model, the coupled-channels 
model\cite{HRK99}, and the resonating group method (RGM) \cite{W37}
are categorized to this 
approach. In this approach, the constant reduced mass is usually used 
for the moment of inertia for the relative motion. 

\item Adiabatic approach

This is in the opposite limit to the sudden approach, and is justified 
if the reaction takes place very slowly. The liquid-drop + shell correction 
model \cite{UGR81} and the adiabatic time-dependent Hartree-Fock (ATDHF) 
method [26--30]
are in this category. 

\end{itemize}

For the region where the colliding nuclei do not significantly overlap with 
each other, both the sudden and the adiabatic approaches would lead to 
a similar result. However, for the region inside the Coulomb barrier, 
one may obtain 
considerably different results depending on which approach one employs. 
Since one would not know a priori which approach is more reasonable, one 
has to investigate both of the two approaches simultaneously. An important 
thing is that, in either of sudden or adiabatic approach, we are now at 
the stage where the present nuclear reaction models have to 
be re-examined by taking into consideration the many-particle nature 
of nuclear reactions. 

\subsection{Sudden approach: role of antisymmetrization}

A typical model in the sudden approach is the double-folding model 
\cite{broglia91}, where the internucleus potential is constructed by 
convoluting an effective nucleon-nucleon interaction 
with the ground state density distributions of the projectile and 
the target nuclei. This potential corresponds to 
the direct (``Hartree'') part of a microscopic ion-ion potential. 
This model will thus be reasonable for the small overlap region, but 
is questionable for the large overlap due to the Pauli exclusion 
principle. 

This can be easily understood if one considers a simple two nucleon system, 
where two nucleons are confined in potential wells whose center is located 
at $\pm R/2$, respectively. If one ignores the Pauli principle and takes 
only the direct term, the 
two particle wave function is given by 
\begin{equation}
\Psi(\vec{r}_1,\vec{r}_2)=\phi_R(\vec{r}_1)\phi_L(\vec{r}_2),
\end{equation}
where $\phi_R$ and $\phi_L$ are the ground state wave function for 
the right and the left hands side of potential wells, respectively. 
From this wave function, the one-body density reads, 
\begin{equation}
\rho(\vec{r})=\langle \Psi|
\delta(\vec{r}-\vec{r}_1)+\delta(\vec{r}-\vec{r}_2)|\Psi\rangle 
=|\phi_R(\vec{r})|^2+|\phi_L(\vec{r})|^2\equiv 
\rho_R(\vec{r})+\rho_L(\vec{r}). 
\label{rho1}
\end{equation}
This is exactly what one obtains in the so called frozen density 
approximation. 
One obtains an unphysically high density matter in the limit of $R\to 0$ 
in this approximation. 
In contrast, if one takes into account the Pauli principle, 
the two particle wave function is given by 
\begin{equation}
\Psi(\vec{r}_1,\vec{r}_2)
=[\phi_R(\vec{r}_1)\phi_L(\vec{r}_2)
-\phi_R(\vec{r}_2)\phi_L(\vec{r}_1)]/\sqrt{2(1-S(R)^2)}, 
\end{equation}
from which one obtains 
\begin{equation}
\rho(\vec{r})=
|\phi_+(\vec{r})|^2+|\phi_-(\vec{r})|^2. 
\label{rho2}
\end{equation}
Here, 
$S(R)=\langle \phi_R|\phi_L\rangle$ is the overlap integral 
between the ``right'' and ``left'' wave functions, and 
$\phi_\pm\equiv (\phi_R\pm\phi_L)/\sqrt{2(1\pm S(R))}$. 
Although the density (\ref{rho2}) is reduced to Eq. (\ref{rho1}) when 
the overlap integral $S(R)$ is small, the two densities are 
considerably different 
if the overlap is large. In fact, the density given 
by Eq. (\ref{rho2}) leads to the ground state density of the two particle 
system in a unified single potential well 
in a natural way 
in the limit of vanishing $R$, 
rather than the unphysical high density matter. 
An important fact is that the Pauli principle plays an essential role 
even in the {\it frozen configuration} approximation. 

In the double folding model, the effect of Pauli principle is partly 
taken into account in a specific manner, {\it i.e.,} 
through the so called knock-on exchange potential, 
where the interacting pair of nucleons are exchanged to each other. 
There are however many other exchange terms from a microscopic 
point of view, which are not included in the double folding model. 
These are the one nucleon exchanges other 
than the knock-on exchange, the two nucleon exchanges, 
and so on. These exchange effects would have to be taken into account 
if one discusses the internucleus potential inside the Coulomb barrier. 

All the exchange effects can be incorporated in the microscopic 
RGM method. A well-known problem of RGM is that it is very difficult to 
apply it to heavy systems. We have recently developed the no-recoil 
approximation, where 
the recoil effect due to the exchange of nucleons between the projectile 
and target nuclei is neglected \cite{HTT06}.  
We have found that the no-recoil approximation works well 
for the $\alpha$ + $^{90}$Zr reaction and heavier systems. 
We expect that this no-recoil approximation will provide a useful 
way to 
apply the RGM method even to heavy systems. 

\subsection{Adiabatic approach: role of coordinate dependent mass}

The adiabatic approximation is valid when the reaction takes place 
so slowly that the non-collective motion adiabatically follows 
the relative motion at every instant. This corresponds to the case where the 
excitation energies for the non-collective degrees of freedom are 
large \cite{THAB94,HTB97}. 
In the context of heavy-ion collisions, the adiabatic approximation 
involves the dynamical change of the density of 
the colliding nuclei. 
At energies around the Coulomb barrier, 
there are good reasons why the adiabatic approach may be reasonable. 
Firstly, the relative velocity between the colliding nuclei is small 
around the barrier. Secondly, the dynamical deformation of the densities 
involves the excitations (on the sudden basis) of the non-collective 
degrees of freedom. Among them, low-lying collective motions can be explicitly 
included in the coupled-channels framework, but the remaining excitations 
are high-lying and the adiabatic treatment should be adequate. 

One of the important consequences of the adiabatic treatment is that 
the moments of inertia become coordinate dependent\cite{THAB94}.  
That is, the Hamiltonian for the relative motion is now given as
\begin{equation}
H=-\frac{\hbar^2}{2}\,\frac{d}{dr}\,\frac{1}{M(r)}\,\frac{d}{dr}+V(r)
+\frac{l(l+1)\hbar^2}{2\Theta(r)}, 
\label{Had}
\end{equation}
where $M(r)$ and $\Theta(r)$ are the moment of inertia for the translational 
and the rotational motions, respectively. 
(There is an additional complication originating from the way of quantization 
with the coordinate dependent mass, 
which we do not discuss in this contribution.)

Using a coordinate transformation, 
the Hamiltonian (\ref{Had}) can be written in a different form as well. 
Transforming the coordinate from $r$ to $\rho$ such that 
$M(r)\to \widetilde{M}(\rho)=\mu$ and 
$\Theta(r)\to \widetilde{\Theta}(\rho)$\cite{UGR81}, 
one obtains
\begin{equation}
H=
-\frac{\hbar^2}{2\mu}\,\frac{d^2}{d\rho^2}+\widetilde{V}(\rho)
+\frac{l(l+1)\hbar^2}{2\widetilde{\Theta}(\rho)}. 
\end{equation}
If one writes this Hamiltonian in a form of
\begin{equation}
H=-\frac{\hbar^2}{2\mu}\,\frac{d^2}{d\rho^2}+
\frac{l(l+1)\hbar^2}{2\mu\rho^2} 
+\widetilde{V}(\rho)
+\frac{l(l+1)\hbar^2}{2\widetilde{\Theta}(\rho)}
-\frac{l(l+1)\hbar^2}{2\mu\rho^2},
\end{equation}
it implies that the internucleus potential has a strong 
angular momentum dependence. 

So far, the possibility of coordinate dependent mass has not 
yet been considered seriously in the context 
of heavy-ion reactions, except for Ref. \cite{M72}.
It would be an interesting future problem to explore it more 
systematically in connection to sub-barrier fusion reactions. 

\begin{theacknowledgments}
This work is based on collaborations with M. Dasgupta, T. Takehi, 
A.B. Balantekin, and N. Takigawa. 
We also thank D.J. Hinde and A. Diaz-Torres for useful discussions. 
This work was supported by the Grant-in-Aid for Scientific Research,
Contract No. 16740139 from the Japanese Ministry of Education,
Culture, Sports, Science, and Technology.
\end{theacknowledgments}


\begin{thebibliography}{99}

\bibitem{RKL89}N. Rowley, A. Kabir, and R. Lindsay, J. Phys. {\bf G15}, 
L269 (1989). 

\bibitem{RM91}N. Rowley and A.C. Merchant, Astrophys. J. {\bf 381}, 
591 (1991). 

\bibitem{RLWL93}N. Rowley, J.R. Leigh, J.X. Wei, and R. Lindsay, 
Phys. Lett. {\bf B314}, 179 (1993). 

\bibitem{L95}J.R. Leigh {\it et al.}, Phys. Rev. C{\bf 52}, 3151 (1995). 

\bibitem{HKT98}K. Hagino, S. Kuyucak, and N. Takigawa, 
Phys. Rev. C{\bf 57}, 1349 (1998). 

\bibitem{HDGHMN01}K. Hagino {\it et al.}, 
in proc. of the 4th Italy-Japan symposium 
on Heavy-Ion Physics, edited by S. Kubono {it et al.} (World Scientific, 
Singapore, 2002), p.87. e-print: nucl-th/0110065. 

\bibitem{HRD03}K. Hagino, N. Rowley, and M. Dasgupta, 
Phys. Rev. C{\bf 67}, 054603 (2003). 

\bibitem{newton04}J.O. Newton {\it et al.}, 
Phys. Lett. B{\bf 586}, 219 
(2004); Phys. Rev. C {\bf 70}, 024605 (2004). 


\bibitem{broglia91}R.A. Broglia and A. Winther, {\it Heavy Ion Reactions}, 
Vol. 84 in Frontiers in Physics Lecture Note Series (Addison-Wesley,
Redwood City, CA, 1991). 

\bibitem{Christensen76}P.R. Christensen and A. Winther, Phys. Lett. {\bf 65B}, 
19 (1976).

\bibitem{SL79}G.R. Satchler and W.G. Love, Phys. Rep. {\bf 55}, 183
  (1979). 

\bibitem{WHD06}K. Washiyama, K. Hagino, and M. Dasgupta, 
Phys. Rev. C{\bf 73}, 034607 (2006). 

\bibitem{hagino05} K. Hagino, T. Takehi, A. B. Balantekin, 
and N. Takigawa, Phys. Rev. C {\bf 71}, 044612 (2005).

\bibitem{hagino04}K. Hagino and N. Rowley, Phys. Rev. C {\bf 69}, 
054610 (2004). 

\bibitem{ARN88}M.V. Andres, N. Rowley, and M.A. Nagarajan, 
Phys. Lett. B{\bf 202}, 292 (1988). 

\bibitem{nanda98}
M.~Dasgupta, D.~J. Hinde, N.~Rowley, and A.~M. Stefanini,
\newblock Ann. Rev. Nucl. Part. Sci. {\bf 48}, 401 (1998).

\bibitem{BT98}
A.B. Balantekin and N. Takigawa,
Rev. Mod. Phys. {\bf 70}, 77 (1998).

\bibitem{rowley91} N. Rowley, G. R. Satchler, and P. H. Stelson,
Phys. Lett. {\bf B254}, 25 (1991).

\bibitem{timmers95}H. Timmers {\it et al.}, 
Nucl. Phys. {\bf A584}, 190 (1995). 

\bibitem{LW81}S. Landowne and H.H. Wolter, Nucl. Phys. {\bf A351}, 171
  (1981).



\bibitem{HD06}D.J. Hinde and M. Dasgupta, private communication. 

\bibitem{HRK99}K. Hagino, N. Rowley, and A.T. Kruppa,
Comp. Phys. Comm. {\bf 123}, 143 (1999).

\bibitem{W37}J.A. Wheeler, Phys. Rev. {\bf 52}, 1083 (1937); 
{\it ibid.}, 1107 (1937). 

\bibitem{UGR81}J.N. Urbano, K. Goeke, and P.-G. Reinhard, 
Nucl. Phys. {\bf A370}, 329 (1981). 

\bibitem{BV78}M. Baranger and M. Veneroni, Ann. of Phys. (N.Y.) 
{\bf 114}, 123 (1978).

\bibitem{BGV76}D.M. Brink, M.J. Giannoni, and 
M. Veneroni, Nucl. Phys. {\bf A258}, 237 (1976). 

\bibitem{FHV80}H. Flocard, P.H. Heenen, and D. Vautherin, 
Nucl. Phys. {\bf A339}, 336 (1980). 

\bibitem{HFV83}P.H. Heenen, H. Flocard, and D. Vautherin, 
 Nucl. Phys. {\bf A349}, 525 (1983). 

\bibitem{H81}P.H. Heenen, Phys. Lett. {\bf 99B}, 298 (1981). 

\bibitem{RFGGG84}P.-G. Reinhard, J. Friedrich, K. Goeke, 
F. Gr\"ummer, and D.H.E. Gross, Phys. Rev. C{\bf 30}, 878 (1984). 

\bibitem{HTT06}K. Hagino, T. Takehi, and N. Takigawa, 
e-print:nucl-th/0603027. 

\bibitem{THAB94}N. Takigawa, K. Hagino, M. Abe and A.B. Balantekin, 
Phys. Rev. C{\bf 49}, 2630 (1994). 

\bibitem{HTB97}K. Hagino, N. Takigawa, and A.B. Balantekin, 
Phys. Rev. C{\bf 56}, 2104 (1997). 

\bibitem{M72}U. Mosel, Particles and Nuclei, {\bf 3}, 297 (1972). 

\end{thebibliography}
\end{document}